\documentclass[twocolumn]{aastex6} 

\usepackage{lineno}

\newcommand{\ind}[1]{_{\mathrm{#1}}}

\shorttitle{Solar Oscillations on Neptune}
\shortauthors{Gaulme et al.}

\begin{document}


\title{A Distant Mirror: Solar Oscillations Observed on Neptune by the \textit{Kepler} K2 Mission}


\author{P. Gaulme\altaffilmark{1,2,3}}
\author{J. F. Rowe\altaffilmark{4}}
\author{T. R. Bedding\altaffilmark{5,11}}
\author{O. Benomar\altaffilmark{6}}
\author{E. Corsaro\altaffilmark{7,8,9,10}}
\author{G. R. Davies\altaffilmark{10,11}}
\author{S. J. Hale\altaffilmark{11,12}}
\author{R. Howe\altaffilmark{11,12}}
\author{R. A. Garcia\altaffilmark{7}}
\author{D. Huber\altaffilmark{5,12,13}}
\author{A. Jim\'enez\altaffilmark{8,9}}
\author{S. Mathur\altaffilmark{14}}
\author{B. Mosser\altaffilmark{15}}
\author{T. Appourchaux\altaffilmark{16}} 
\author{P. Boumier\altaffilmark{16}}
\author{J. Jackiewicz\altaffilmark{1}}
\author{J. Leibacher\altaffilmark{16,17}}
\author{F.-X. Schmider\altaffilmark{18}}
\author{H. B. Hammel\altaffilmark{19}}
\author{J. J. Lissauer\altaffilmark{20}}
\author{M. S. Marley\altaffilmark{20}}
\author{A. A. Simon\altaffilmark{21}}
\author{W. J. Chaplin\altaffilmark{11,12}}
\author{Y. Elsworth\altaffilmark{11,12}}
\author{J. A. Guzik\altaffilmark{22}}
\author{N. Murphy\altaffilmark{23}}
\author{V. Silva Aguirre\altaffilmark{12}}

\altaffiltext{1}{Department of Astronomy, New Mexico State University, P.O. Box 30001, MSC 4500, Las Cruces, NM 88003-8001, USA}
\altaffiltext{2}{Apache Point Observatory, 2001 Apache Point Road, P.O. Box 59, Sunspot, NM 88349, USA}
\altaffiltext{3}{Physics Department, New Mexico Institute of Mining and Technology, 801 Leroy Place, Socorro, NM 87801, USA}
\altaffiltext{4}{Institut de recherche sur les exoplan\`etes, iREx, D\'epartement de physique, Universit\'e de Montr\'eal, Montr\'eal, QC, H3C 3J7, Canada}
\altaffiltext{5}{Sydney Institute for Astronomy (SIfA), School of Physics, University of Sydney, NSW 2006, Australia}
\altaffiltext{6}{Center for Space Science, NYUAD Institute, New York University Abu Dhabi, PO Box 129188, Abu Dhabi, UAE}
\altaffiltext{7}{Laboratoire AIM, CEA/DRF-CNRS, Universit\'e Paris 7 Diderot, IRFU/SAp, Centre de Saclay, 91191 Gif-sur-Yvette, France}
\altaffiltext{8}{Instituto de Astrof\'{\i}sica de Canarias, E-38200 La Laguna, Tenerife, Spain}
\altaffiltext{9}{Departamento de Astrof\'{\i}sica, Universidad de La Laguna, E-38205 La Laguna, Tenerife, Spain}
\altaffiltext{10}{INAF - Osservatorio Astrofisico di Catania, Via S. Sofia 78, I-95123 Catania, Italy}
\altaffiltext{11}{School of Physics \& Astronomy, University of Birmingham, Edgbaston, Birmingham, B152TT, UK}
\altaffiltext{12}{Stellar Astrophysics Centre (SAC), Department of Physics and Astronomy, Aarhus University, Ny Munkegade 120, DK-8000 Aarhus C, Denmark}
\altaffiltext{13}{SETI Institute, 189 Bernardo Avenue, Mountain View, CA 94043, USA} 
\altaffiltext{14}{Center for Extrasolar Planetary Systems, Space Science Institute, 4750 Walnut street, Suite\#205, Boulder, CO 80301, USA}
\altaffiltext{15}{LESIA, Observatoire de Paris, PSL Research University, CNRS, Universit\'e Pierre et Marie Curie, Universit\'e Denis Diderot, 92195 Meudon, France}
\altaffiltext{16}{Institut d'Astrophysique Spatiale, Universit\'e Paris-Sud and CNRS (UMR 8617), B\^{a}timent 121, F-91405 Orsay cedex, France}
\altaffiltext{17}{National Solar Observatory, 950 N. Cherry Ave., Tucson, AZ 85718, USA}
\altaffiltext{18}{Laboratoire Lagrange, Observatoire de la C\^ote d'Azur, Universit\'e de Nice-Sophia-Antipolis, CNRS, Nice, France}
\altaffiltext{19}{AURA, Inc., 1331 Pennsylvania Avenue NW, Suite 1475, Washington, DC 20004, USA}
\altaffiltext{20}{NASA Ames Research Center, Space Science \& Astrobiology Division, MS 245-3, Moffett Field, CA 94035, USA}
\altaffiltext{21}{NASA Goddard Space Flight Center, Solar System Exploration Division (690.0), 8800 Greenbelt Road, Greenbelt, MD, 20771, USA}
\altaffiltext{22}{Los Alamos National Laboratory, XTD-NTA, MS T086, Los Alamos, NM 87545-2345, USA}
\altaffiltext{23}{Jet Propulsion Laboratory / Caltech, 4800 Oak Grove Drive, Pasadena CA 91109}

\email{gaulme@nmsu.edu}



\begin{abstract}
Starting in December 2014, \textit{Kepler} K2 observed Neptune continuously for 49 days at a 1-minute cadence. The goals consisted of studying its atmospheric dynamics \citep{Simon_2016}, detecting its global acoustic oscillations (Rowe et al., submitted), and those of the Sun, which we report on here. We present the first indirect detection of solar oscillations in intensity measurements. Beyond the remarkable technical performance, it indicates how \textit{Kepler} would see a star like the Sun. The result from the global asteroseismic approach, which consists of measuring the oscillation frequency at maximum amplitude $\nu\ind{max}$ and the mean frequency separation between mode overtones $\Delta\nu$, is surprising as the $\nu\ind{max}$ measured from Neptune photometry is larger than the accepted value. Compared to the usual reference $\nu\ind{max,\odot} = 3100\ \mu$Hz, the asteroseismic scaling relations therefore make the solar mass and radius appear larger by $13.8\pm 5.8\,\%$ and $4.3\pm1.9\,\%$ respectively. The higher $\nu\ind{max}$ is caused by a combination of the value of $\nu\ind{max,\odot}$, being larger at the time of observations than the usual reference from SOHO/VIRGO/SPM data ($3160\pm10\ \mu$Hz), and the noise level of the K2 time series, being ten times larger than VIRGO's. The {\it peak-bagging} method provides more consistent results: despite a low signal-to-noise ratio (SNR), we model ten overtones for degrees $\ell=0,1,2$. We compare the K2 data with simultaneous SOHO/VIRGO/SPM photometry and BiSON velocity measurements. The individual frequencies, widths, and amplitudes mostly match those from VIRGO and BiSON within $1 \sigma$, except for the few peaks with lowest SNR.
\end{abstract}



\section{The importance of reflected solar modes}
\label{sect_intro}
It is well known that the Sun exhibits oscillations on a 5-minute timescale due to convection-driven pressure modes.
The Sun's disk-integrated helioseismic properties are a standard reference which has become increasingly relevant due to asteroseismic information that can be routinely extracted from high-quality, high-cadence, long duration time-series provided by missions such as CoRoT and {\it Kepler} \citep{Baglin_2009,Borucki_2010}.

Ideally, the measurements of solar oscillations that act as a reference should be observed with the same
instrument as the stars.  
Observations of Neptune with K2 allowed for a unique opportunity to measure integrated disk seismic properties of the Sun in reflected light and determine fundamental properties (mass, radius) of the Sun as a distant star. 
Solar oscillations have been measured in radial velocity from the Moon \citep{Fussell_1995,Kjeldsen_2005}, and also from the blue sky in both equivalent width \citep{Kjeldsen_1995} and radial velocity \citep{Kjeldsen_2008}. To our knowledge, our analysis of K2 photometric observations of reflected solar light from Neptune is the first indirect detection of solar oscillations in intensity. 

To first approximation, the Fourier spectrum of solar-like oscillations consists of a series of overtone modes that are regularly spaced in frequency with a separation of $\Delta\nu$, under a broad envelope that is centered at $\nu\ind{max}$.  The solar values for these quantities are approximately $\Delta\nu_\sun = 134.9\ \mu$Hz and $\nu\ind{max,\sun} = 3100\ \mu$Hz \citep[e.g.][]{Broomhall_2009}.  Theoretical calculations have established that, to a good approximation, $\Delta\nu$ is proportional to the square root of the mean stellar density \citep[e.g.][]{Ulrich_1986}.  The scaling of $\nu\ind{max}$ to other stars, on the other hand, is less secure. \citet{Brown_1991} conjectured that $\nu\ind{max}$ should scale as $g/\sqrt{\ind{Teff}}$, and this has been used to predict the properties of oscillations in other stars \citep{Kjeldsen_Bedding_1995}.
Subsequently, \citet{Stello_2008} suggested that the observed value of $\nu\ind{max}$ could be used to infer relative stellar properties. 
It has become common to determine stellar properties, such as a mass $M$ and radius $R$, from measurements of $\nu\ind{max}$, $\Delta\nu$ and the effective temperature ($T\ind{eff}$) relative to the Sun:  

\begin{eqnarray}
\label{eq_scaling_M}
\frac{M}{M_\odot}\ &=&\  \left(\frac{\nu\ind{max}}{\nu\ind{max_\odot}}\right)^3\ \left(\frac{\Delta\nu_\odot}{\Delta\nu}\right)^4\ \left(\frac{T\ind{eff}}{T\ind{eff,\odot}}\right)^{\frac{3}{2}}\\
\frac{R}{R_\odot}\ &=&\ \left(\frac{\nu\ind{max}}{\nu\ind{max_\odot}}\right)\ \ \left(\frac{\Delta\nu_\odot}{\Delta\nu}\right)^2\ \left(\frac{T\ind{eff}}{T\ind{eff,\odot}}\right)^{\frac{1}{2}}.
\label{eq_scaling_R}
\end{eqnarray}
These {\it asteroseismic scaling relations} are widely used and appear to be valid over a large range of stellar types that exhibit p-mode oscillations, ranging from main-sequence dwarfs to evolved red giants \citep[see][for recent reviews]{Chaplin_Miglio_2013, Belkacem_2013}.

An important note is that the scaling relation for $\nu\ind{max}$  is largely empirical and the determination of $\nu\ind{max}$  depends on the details of the observations (e.g. photometric bandpass) and the method used to extract $\nu\ind{max}$ from the observed time-series. Indeed, the photometric amplitudes of the modes, including the relative amplitudes of modes with different angular degrees, vary with the wavelength of the optical spectrum \citep{Bedding_1996,Michel_2009}.  The published values of $\nu\ind{max}$ range from 3050 \citep[][]{Kjeldsen_Bedding_1995} to $3150\ \mu$Hz \citep{Chaplin_2011c}. For this reason, it would be very useful to measure $\nu\ind{max}$ in the \textit{Kepler} bandpass.  The K2 observations of Neptune provide this opportunity, although it has to be kept in mind that the albedo of Neptune is a function of wavelength \citep[see][Fig. 4, for comparison of Kepler bandpass and Neptune's atmospheric penetration depth]{Simon_2016}.

Another reason for our interest in the K2 observations of Neptune is to calibrate the {\em amplitudes} of oscillations in the Sun.  There has been considerable effort towards understanding how the amplitudes of solar-like oscillations vary from the Sun to other stars, both theoretically \citep{Christensen_Dalsgaard_Frandsen_1983,Kjeldsen_Bedding_1995, Kjeldsen_Bedding_2011, Houdek_1999, Houdek_Gough_2002, Houdek_2006, Samadi_2007, Samadi_2012,Belkacem_2011b}
and observationally \citep{Samadi_2010, Huber_2011, Huber_2011b, Campante_2011, Chaplin_2011b, Belkacem_2012, Mosser_2012b, Corsaro_2013}.
Once again, a good measurement of the solar amplitude with \textit{Kepler} would
serve as an important calibration.

In this Letter, we report the detection and analysis of the solar oscillation spectrum from photometric measurements of solar light reflected by Neptune. We firstly treat the oscillation spectrum as we would do with any other \textit{Kepler} target, by measuring its global parameters $\Delta\nu$ and $\nu\ind{max}$. Then we model the oscillation spectrum with a standard ``peak-bagging'' approach, to extract individual mode frequencies, widths, and amplitudes for eight orders. We compare these results with Birmingham Solar-Oscillations Network (BiSON) and SOHO/VIRGO/SPM data.

\section{Data and Analysis}
\subsection{From raw images to a clean light curve}
\begin{figure}[t]
\epsscale{1.15}
\plotone{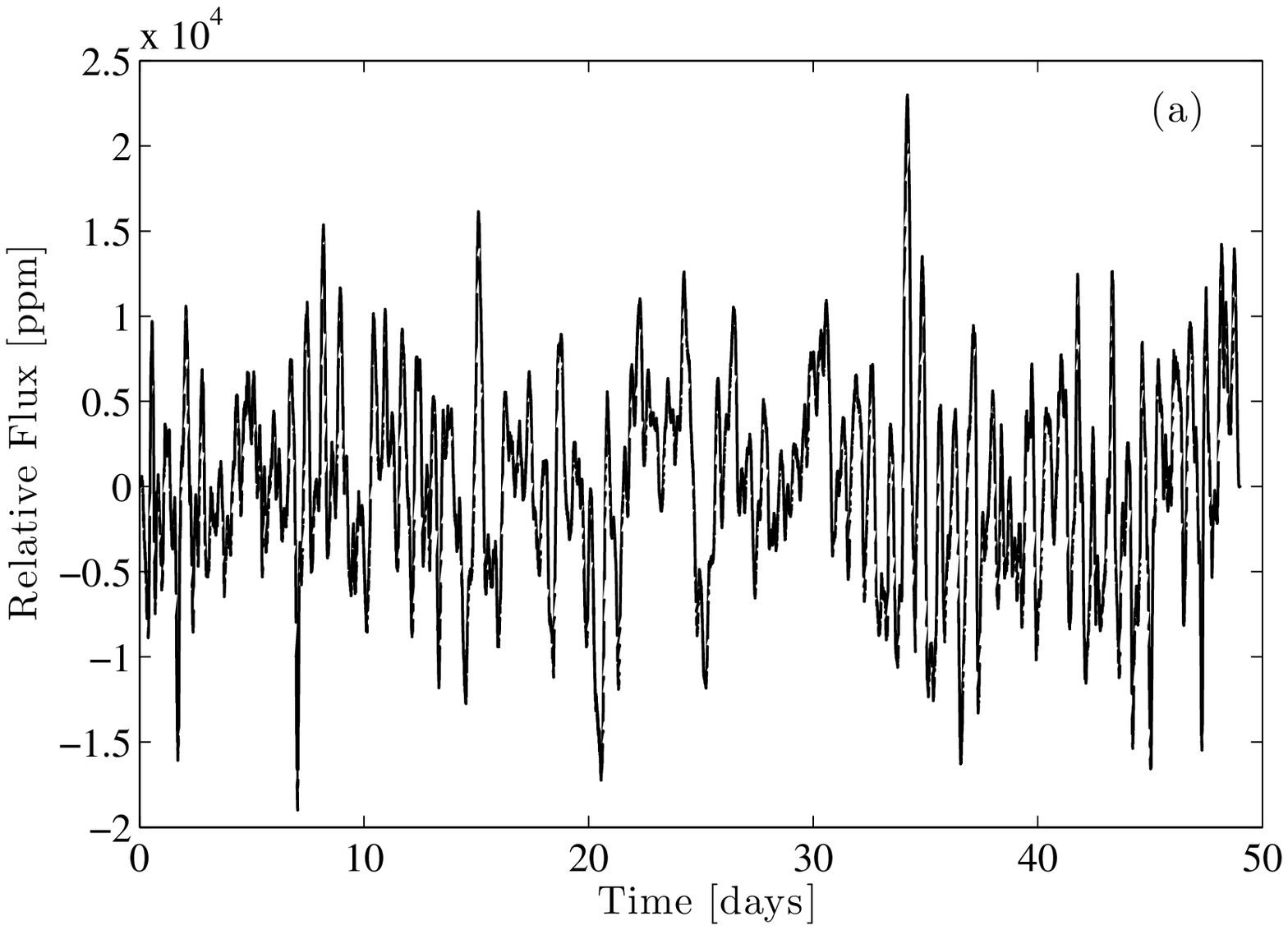}
\plotone{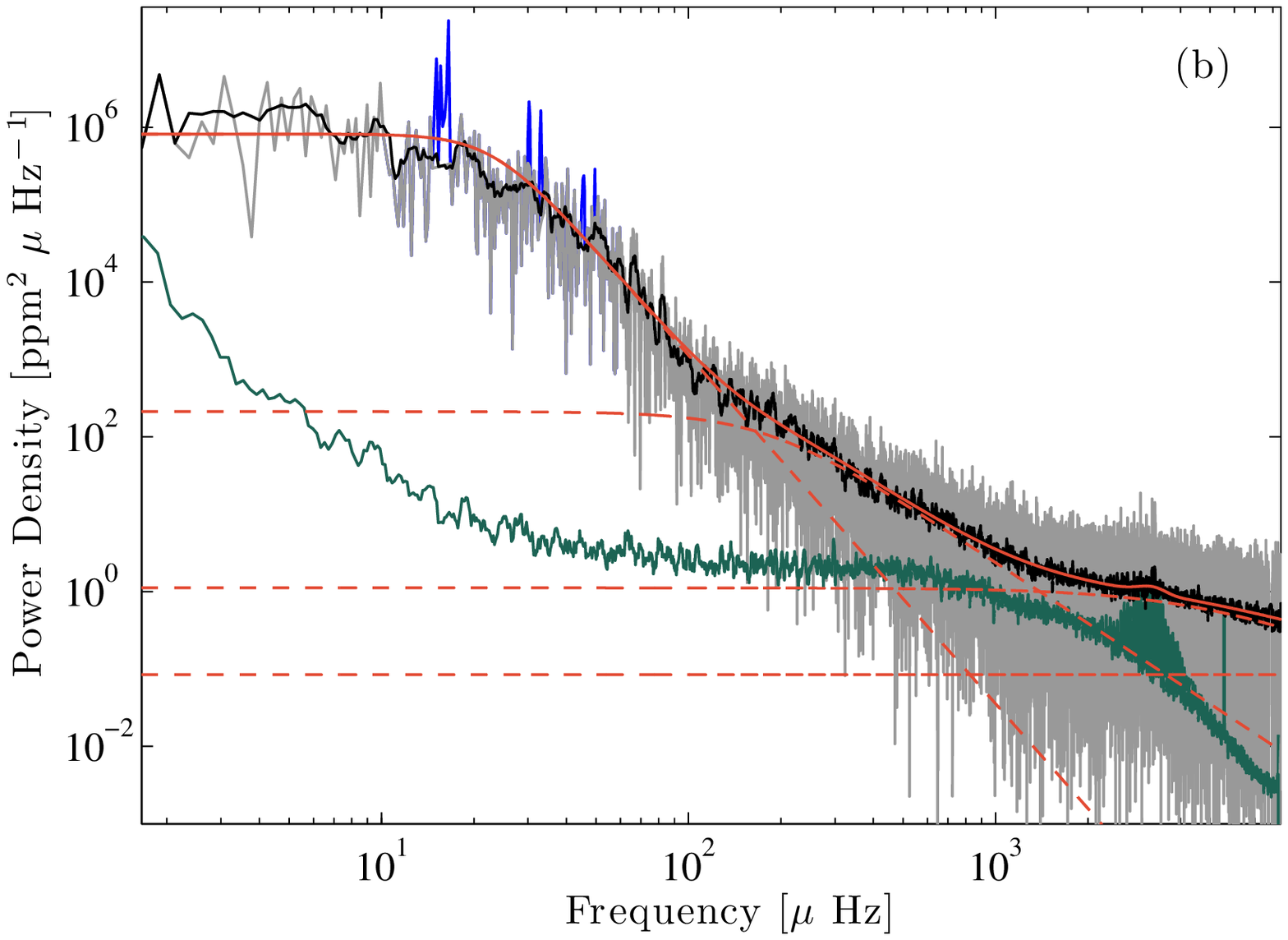}
\caption{Panel a: K2 Neptune full 49-day light curve, showing normalized brightness variations over time elapsed since December 1, 2014. Panel b: gray line is the power density spectrum of the \textit{Kepler} light curve in the square of parts per million (ppm$^2$) per $\mu$Hz, as a function of frequency ($\mu$Hz). Blue peaks are Neptune's rotation frequencies and harmonics. Black line is the power density smoothed over 100 bins to guide the eye to the mean noise level. The plain red line indicates the noise model plus the mode envelope, which is the sum of three semi-Lorentzians, a Gaussian, and a white noise offset (dashed red lines). The excess power due the solar modes is visible in the bottom right of the plot. The green line is the smoothed (100 bins) power density spectrum of the VIRGO/SPM light curve taken simultaneously with K2 data. \label{fig_data}}
\end{figure}

We used the corrected 49-day K2 photometric lightcurve reported by Rowe et al. (submitted), which includes corrections for photometric jumps, intrapixel variations and outliers. The light curve was detrended to remove the observed decrease in flux due to the increasing distance between Neptune and the {\it Kepler} spacecraft, by subtracting a second-order polynomial (Fig. \ref{fig_data}a). Over the 49-day observation window the distance between the {\it Kepler} spacecraft and Neptune increased by 0.81 AU, which represents a 406-second variation of light travel time. Since we consider physical phenomena on the Sun, we interpolated the data onto a uniform time grid that takes into account the light travel time. We also accounted for the distance variation from the Sun to Neptune, even though it is very small (0.8 seconds).

We analyzed the light curve in terms of frequencies by computing its power spectral density (PSD) with a Fast Fourier Transform.  All short gaps (only a few missing points) were interpolated with a second-order polynomial estimated from the nearby data points. There are no long gaps observed and the overall duty cycle is greater than 98\,\%. The power spectrum is single sided and was properly calibrated to satisfy Parseval's theorem \citep[e.g.,][]{Appourchaux_2014}.

\subsection{Extraction of  the Sun's global helioseismic parameters}
The global asteroseismic approach involves measuring  $\Delta\nu$  and $\nu\ind{max}$ from the PSD of the light curve. 

To determine $\nu\ind{max}$, one must model background noise in the frequency domain, which is typically dominated by the correlated stellar noise (spots, granulation, meso- and super-granulation). To determine $\nu\ind{max}$, we fitted the background with a sum of two or three super Lorentzians centered on zero frequency, a Gaussian accounting for the mode envelope, and white noise \citep{Harvey_1985}. The center of the Gaussian constitutes our measurement of $\nu\ind{max}$. The model $S$ is expressed as:
\begin{equation}
S(\nu)\ = \sum_{i} \frac{H_i}{\displaystyle 1 + (\tau_i \nu)^{p_i}}\ +\ \exp\left(\frac{\nu-\nu\ind{max}}{2\sigma^2}\right)+\ B_0,
\label{Eq_harvey}
\end{equation}
where $\nu$ indicates the frequency, $(H_i, \tau_i, p_i)$ are the height, characteristic time and slope of each super-Lorentzian, $\sigma$ is the Gaussian standard deviation, and $B_0$ is the white noise. 

In our case, the background variability arises from sources other than solar spots and granulation. This is obvious when comparing the K2 PSD with simultaneous SOHO/VIRGO (green channel) data (Fig. \ref{fig_data}). K2's background overwhelms VIRGO's by up to four orders of magnitude. Despite the application of optimized techniques for correcting instrumental effects such as intrapixel variability and gain variations \citep{Rowe_2016}, there remains a large noise level. It is unlikely that Neptune atmospheric features are responsible for such a noise level given the planet's smooth aspect in the visible, except for isolated cloud structures that appear as outstanding peaks between 15 and 17 $\mu$Hz, and their harmonics at $[30,33]$ and $[45,50]\ \mu$Hz \citep[e.g.][]{Simon_2016}. Note that we removed these peaks from the PSD when fitting the background noise to not bias the result.
 
To minimize the bias in estimating the global asteroseismic parameters, we measured them independently with seven slightly different approaches, by different groups. The idea was to proceed as we would if this target were one of the many oscillating stars detected by \textit{Kepler}. In other words, we let each group use its own method, which we detail here.

Co-authors Gaulme, Garc\'{i}a/Mathur, and Mosser measured $\Delta\nu$ from the autocorrelation of the time series \citep{Mosser_Appourchaux_2009}, whereas Huber used the autocorrelation of the power spectrum \citep{Huber_2009}. Benomar, Corsaro, and Davies estimated $\Delta\nu$ with very similar approaches, based on a linear fitting of the individual radial mode frequencies of the modes with larger SNR \citep[for details, see][]{Benomar_2012,Corsaro_2013,Davies_2016b}.

Different methods here used for determining $\nu\ind{max}$, with the number of Lorentzians and of free parameters depending on the approach. Co-authors Benomar and  Gaulme considered two and three Lorentzians respectively, with all parameters free, and Bayesian numerical methods described by \citet{Benomar_2012,Gaulme_2009}. In both cases, no priors on the granulation time scales and slopes were imposed, assuming the solar priors were not sensible because the spectrum is dominated by other sources of noise. Co-authors Corsaro and Garc\'{i}a/Mathur adopted three-Lorentzian profiles, and used the Bayesian code DIAMONDS and A2Z, respectively \citep{Corsaro_2014,Mathur_2010a}. 
Co-author Davies used a model including three Lorentzians \citep{Davies_2016b}, and performed the fit using EMCEE \citep{Foreman-Mackey_2013}. 
Co-author Huber applied the SYD pipeline \citep{Huber_2009} using two Harvey profiles and fitted the background between 1000 and $7000\ \mu$Hz but excluding the power excess region. The amplitude and $\nu\ind{max}$ were retrieved from the whitened power spectrum heavily smoothed with a Gaussian with FWHM$ = 4\Delta\nu$ \citep{Kjeldsen_2008}. Uncertainties of all quantities were derived from Monte-Carlo simulations as described by \citet{Huber_2011}. Like Huber, co-author Mosser estimated $\nu\ind{max}$  from the maximum value of the smoothed whitened oscillation spectrum.

\begin{figure}[t]
\epsscale{1.1}
\plotone{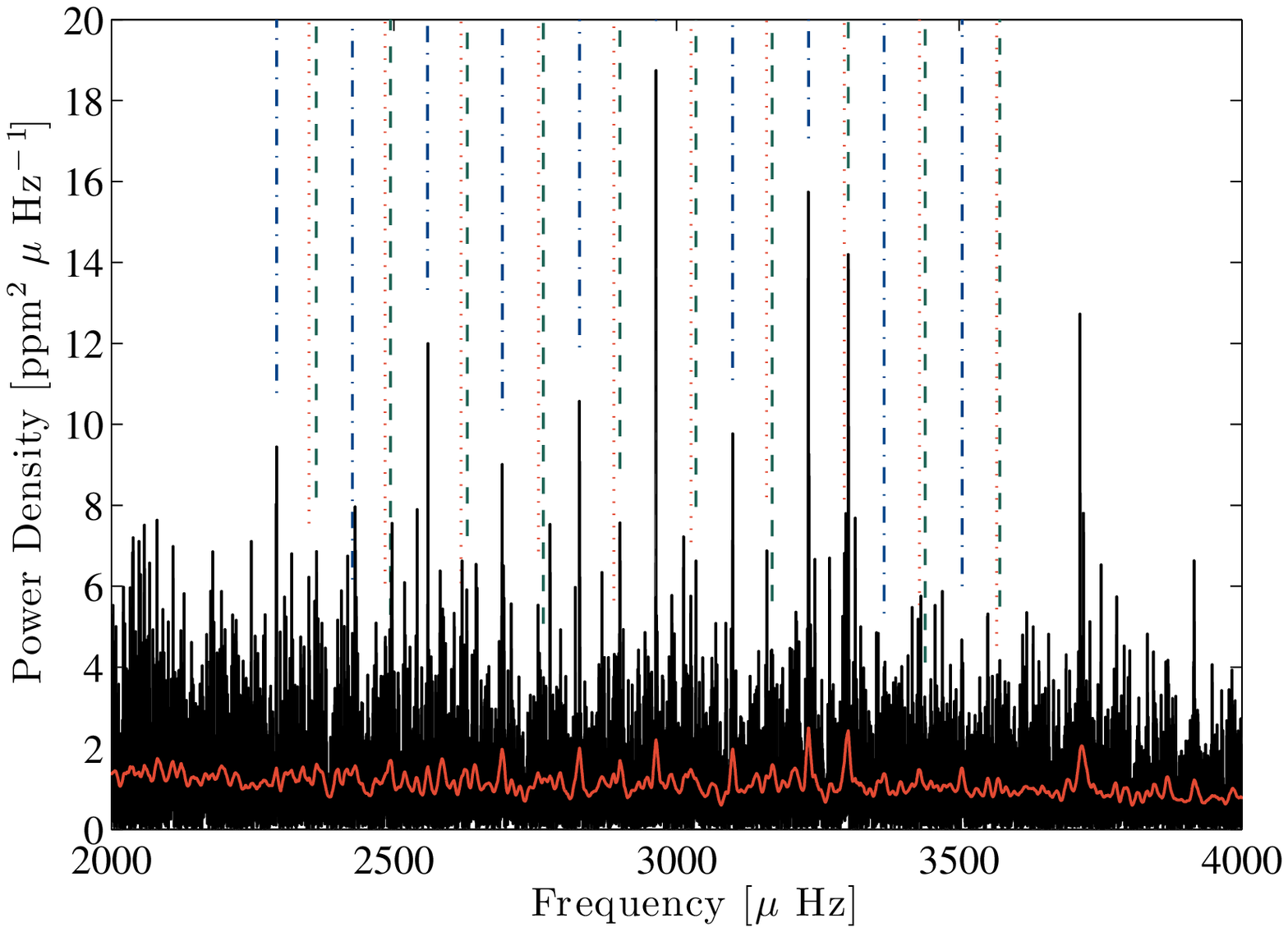}
\plotone{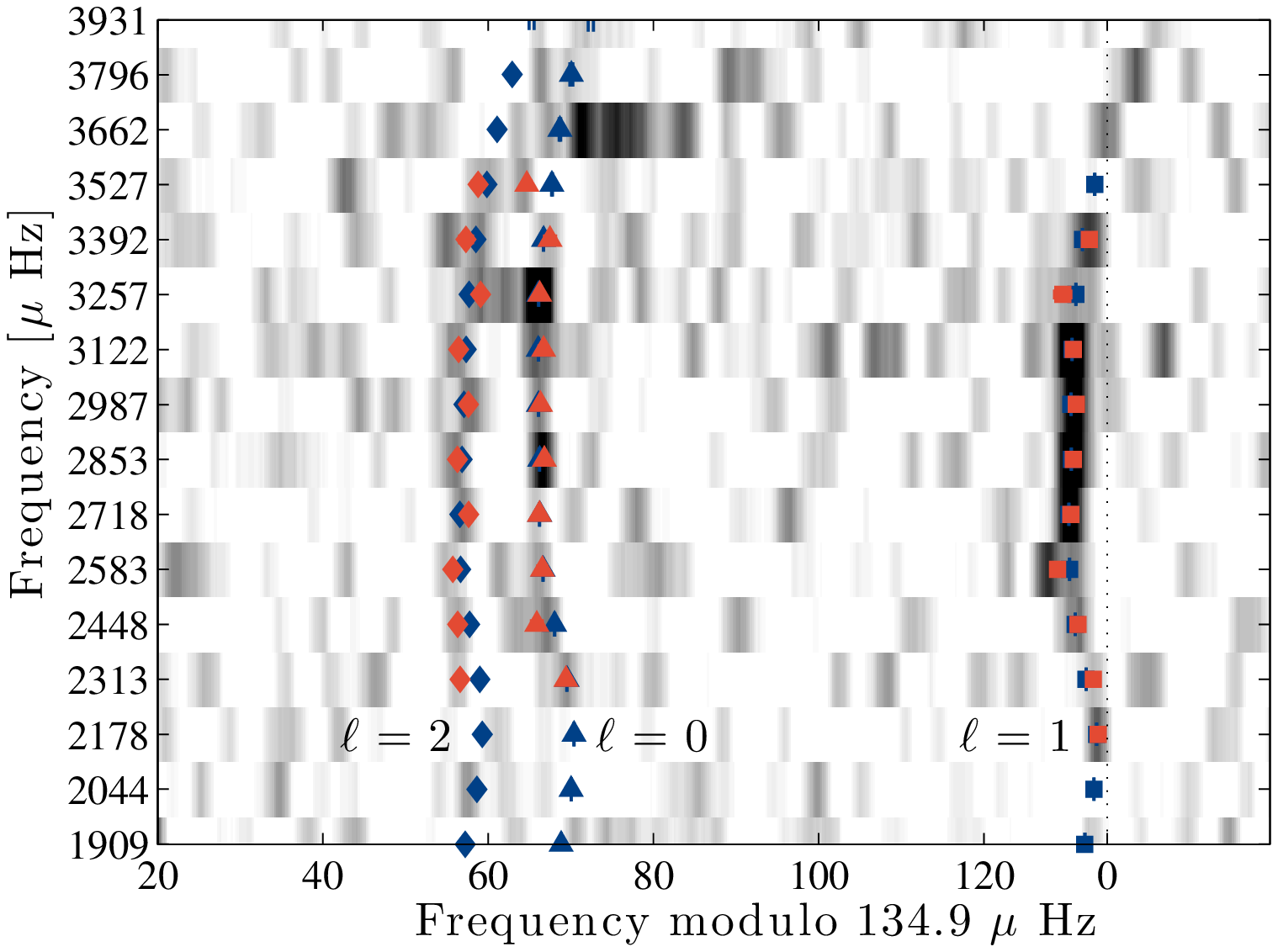}
\caption{Top panel: K2 Neptune power spectral density on a linear scale (ppm$^2\ \mu$Hz$^{-1}$) as a function of frequency ($\mu$Hz). Bottom panel: \'echelle diagram of the power density spectrum. Darker regions correspond to larger peaks in power density. The power density spectrum is smoothed by a weighted moving average over seven bins and folded into $134.9\ \mu$Hz chunks; each is then stacked on top of its lower-frequency neighbor. Red dots indicate K2 frequencies (Corsaro) and the blue dots BiSON's. Error bars are smaller than symbols. \label{fig_ech}}
\end{figure}

\subsection{Extracting individual mode properties}
\label{sect_peakbag}
Modeling an oscillation spectrum, ``peak-bagging'', consists of determining each mode's frequency, height, and width, and possibly also measuring the rotational splitting and rotation axis inclination from the non-radial modes. With a single sided power spectral density, the amplitude of a given mode is defined as $A=\sqrt{H W \pi/2}$ \citep{Appourchaux_2015}, where $H$ and $W$ are its height and width. Fittings were performed by co-authors Benomar, Corsaro, Davies, and Gaulme. To check our results, co-authors Hale and Howe produced and modeled the oscillation spectrum obtained with simultaneous BiSON data, while Corsaro and Gaulme did the same for VIRGO/SPM (green channel). Again, to ensure as much freedom as possible in modeling the data, no specific instruction was given to the fitters. 

In principle, the rotational splitting and the inclination of the rotation axis can be determined from global fitting of an oscillation spectrum \citep[e.g.][]{Gizon_Solanki_2003}. However, given the low SNR of these observations, all fitters fixed the inclination at $90^\circ$ in their final model otherwise the model parameters would not converge properly. As regards the splitting, all co-authors also fixed this parameter, except Benomar, who obtained 0$.45\pm0.22\ \mu$Hz, a result compatible with the actual solar value \citep[$0.434\pm0.002\ \mu$Hz][]{Chaplin_2001}. All performed a global fit of the low-degree modes ($\ell=0, 1,2$), and all modeled eight orders, but Corsaro who considered ten. All fittings, except for BiSON data, were based on a Bayesian approach, i.e. by maximising the likelihood of a model weighted by prior information \citep[e.g.][]{Gregory_2005,Appourchaux_2008a}. 

Benomar 
performed the global fitting with an MCMC algorithm, using a smoothness condition on frequencies \citep{Benomar_2009a,Benomar_2013}. An accurate measure of the mean large frequency spacing was obtained by fitting a linear function to the individual frequencies. 
%
Corsaro performed a peak-bagging analysis with the public code DIAMONDS \citep{Corsaro_2014, Corsaro_2015}. It consisted of a preliminary fit of the background components, and a subsequent fit with a peak significance test, and mode identification. 
Davies used the KAGES procedure \citep{Davies_2016a} for peak bagging, which requires mode identification by inspection, followed by a fit to the data.  After the fit was performed, a machine learnt Bayesian mixture model was fitted to the modes pair-by-pair to estimate the probability that a mode had been detected. 
Gaulme performed a peak-bagging analysis with a maximum \textit{a posteriori} method that associates a maximum likelihood estimator with Bayesian priors. Loose Gaussian priors are applied to mode frequencies, heights and widths, from a smoothing of the power density spectrum \citep{Gaulme_2009}.
Note that Benomar and Corsaro fitted each peak with an independent amplitude, while Davies and Gaulme assumed a uniform amplitude in each order, weighted by mode visibilities. Davies left the visibility factors be free, whereas Gaulme fixed it at $(\ell=1/\ell=0) = 1.5$ and $(\ell=2/\ell=0) = 0.5$. 



\section{Results and Discussion}

\subsection{Solar mass and radius from global parameters}

Measurements of $\nu\ind{max}$ and $\Delta\nu$ based on different methods are presented in Table \ref{tab_1}. 
The measurements of $\nu\ind{max}$ ranged from $3207\pm49$ (Davies) to $3268\pm56\ \mu$Hz (Garc\'{i}a/Mathur) and are consistent within the uncertainties.  The measurements of $\Delta\nu$ ranged from $134.6\pm0.3$ (Davies) to $\Delta\nu=135.3 \pm 0.3\ \mu$Hz (Mosser), again consistent with each other.  

The asteroseismic scaling relations require the effective temperature $T\ind{eff}$, and reference solar values $T\ind{eff,\odot}$, $\Delta\nu_\odot$, and $\nu\ind{max,\odot}$ (Eqs. \ref{eq_scaling_M} \& \ref{eq_scaling_R}). To determine the stellar mass and radius via the asteroseismic scaling relations, we adopted $T\ind{eff\odot} = 5777$~K, $\nu\ind{max,\odot} = 3100\ \mu$Hz and $\Delta\nu_\odot = 134.9\ \mu$Hz.
The mass ranges from $1.11\pm0.05$ (Benomar) to $1.16\pm0.09\ M_\odot$ (Garc\'{i}a/Mathur) and the radius from $1.04\pm0.02$ to $1.05\pm0.03\ R_\odot$. Overall the mass and radius are overestimated on average by about $13.8\pm 5.8\,\%$ and $4.3\pm1.9\,\%$ respectively, i.e. they are off by a little more than 2 $\sigma$. 

\begin{figure}[t]
\epsscale{1.15}
\plotone{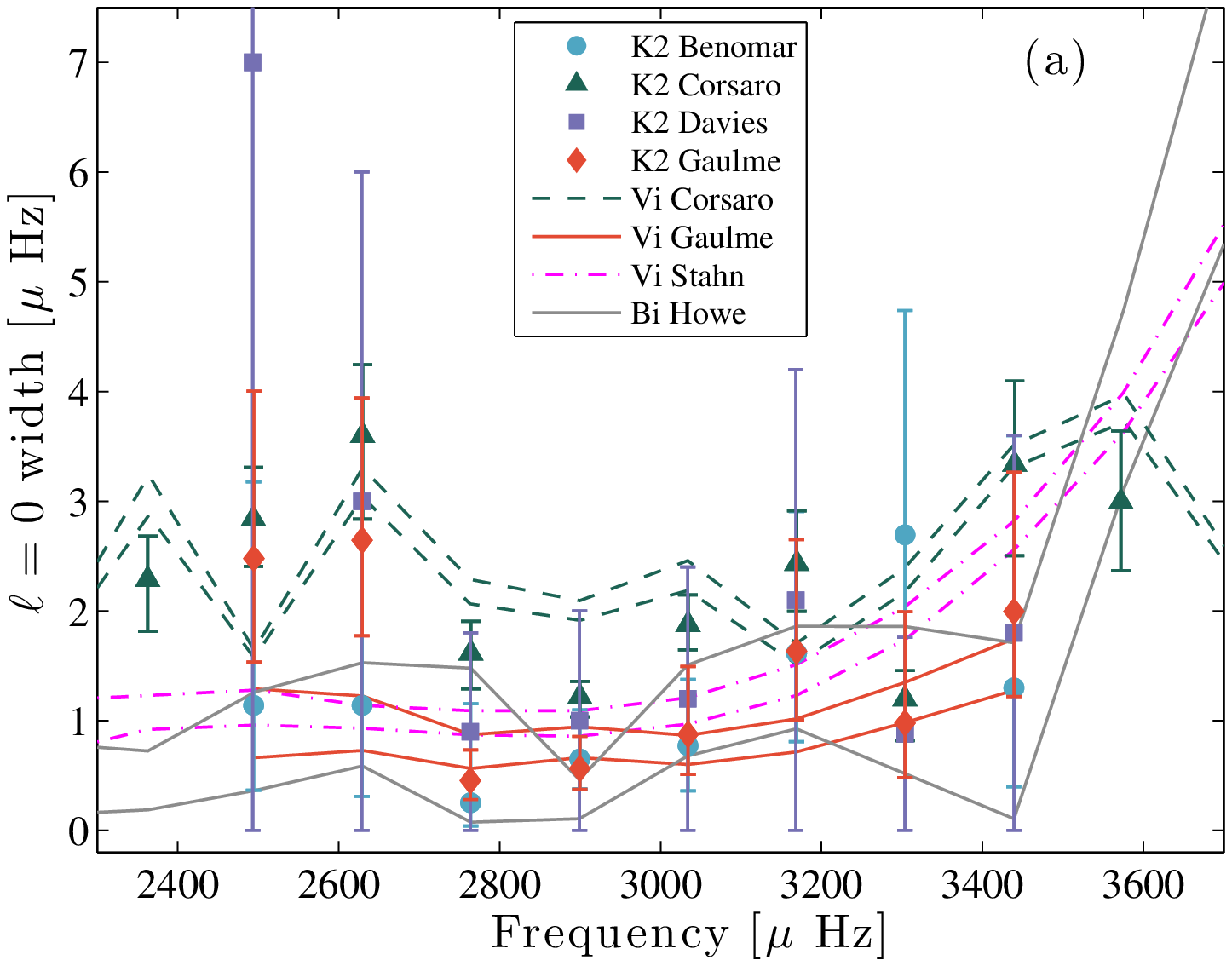}
\plotone{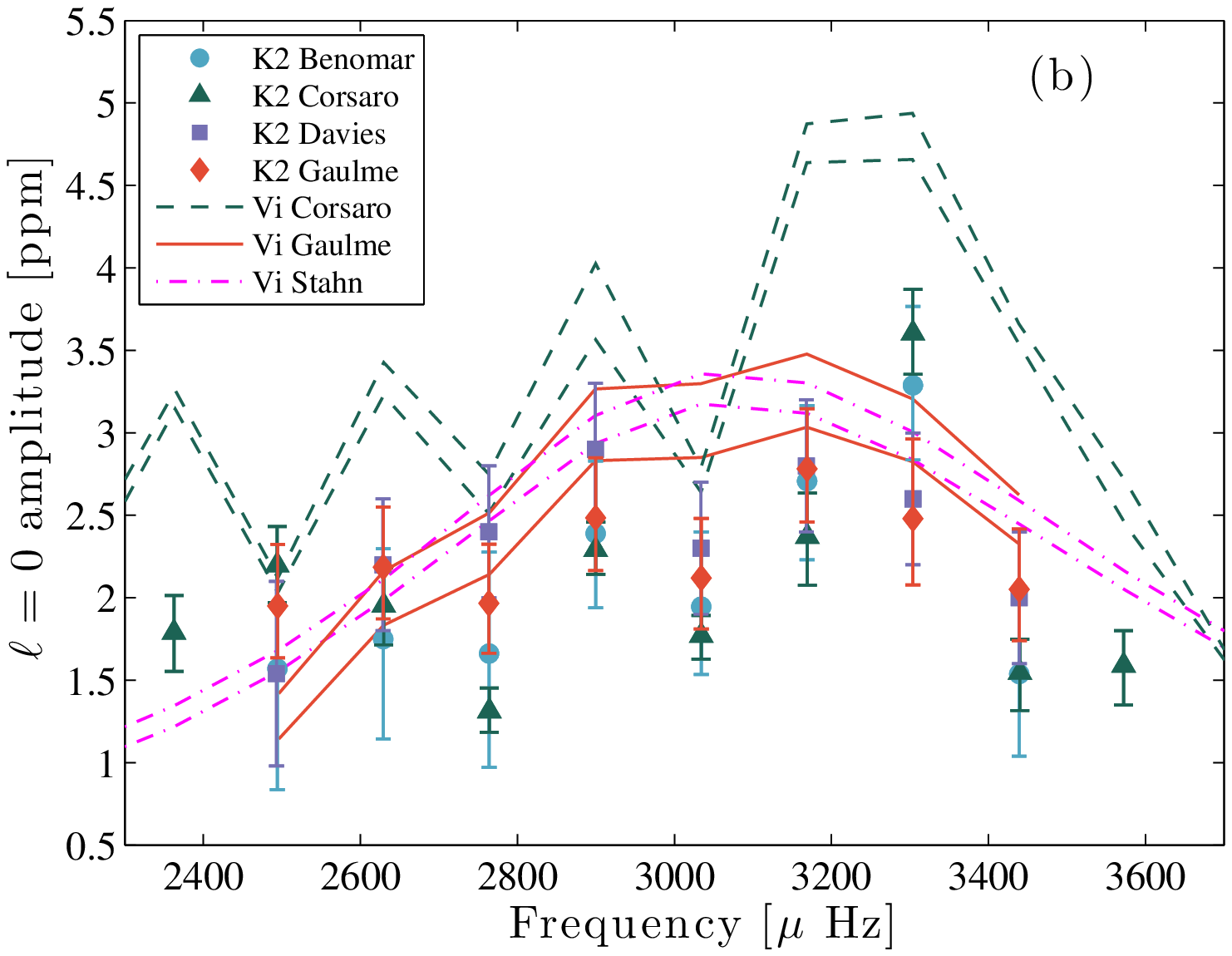}
\caption{Oscillation width and amplitude as a function of frequency for radial modes. Dashed and dash-dotted lines indicate estimates from VIRGO/SPM/green data (black for simultaneous and magenta for 14 years data).\label{fig_W_amp}}
\end{figure}

\begin{figure}[t]
\epsscale{1.15}
\plotone{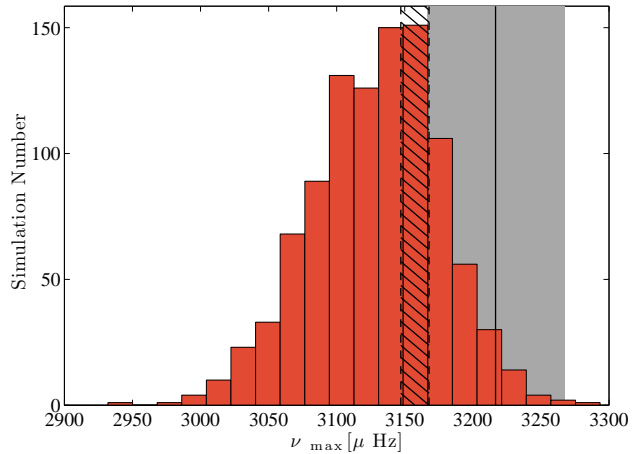}
\caption{Measurement of $\nu\ind{max}$ from VIRGO/SPM (green channel) data, taken simultaneously to K2's, and artificially noised at the K2 level. The histogram is the result of 1000 Monte-Carlo simulations. The gray and hatched areas correspond with Gaulme's $\nu\ind{max}$ from the K2 and the original simultaneous VIRGO data respectively. \label{fig_histo}}
\end{figure}

At first glance, this result is surprising because the ensemble asteroseismic approach is commonly considered to be simple, quick, and reliable. The disagreement can be caused by the actual solar $\nu\ind{max}$ at the time of K2 observations, which fluctuates because of the stochastic nature of the oscillations, and by the large noise level in the K2 PSD, which is ten times larger at $\nu\ind{max}$ than in VIRGO/SPM (green) data. We first checked the simultaneous VIRGO data, and we found $\nu\ind{max} = 3163\pm7$  (Corsaro) and $3158\pm10\ \mu$Hz (Gaulme), which is larger by about $60\ \mu$Hz than the usual $\nu\ind{max,\odot}$, whereas $\Delta\nu = 134.82\pm0.08$ (Corsaro) and $134.65\pm0.28\ \mu$Hz (Gaulme) is consistent with the accepted reference. 
To study the impact of noise on $\nu\ind{max}$, we contaminated the VIRGO light curve with random noise of mean level corresponding to the mean K2 background noise. Gaulme ran 1000 simulations with new random noise at each iteration, and measured $\nu\ind{max}$ each time (Fig. \ref{fig_histo}). 
The mode of the distribution peaks at about $3150\ \mu$Hz. The posterior density distribution, approximated by the histogram, shows that finding $\nu\ind{max}\geq3160\ \mu$Hz has a 20\,\% chance of happening. Thus, by considering the actual solar reference from simultaneous VIRGO photometric measurements $\nu\ind{max,\odot} \approx 3160\pm10\ \mu$Hz, the K2 data lead to masses from $M = 1.05\pm0.05\ M_\odot$ (Benomar) to $1.10\pm0.08\ M_\odot$ (Garc\'{i}a/Mathur), and radii from $R=1.02\pm0.02\ R_\odot$ to $1.03\pm0.03\ R_\odot$, which are within 1 $\sigma$.


\subsection{Individual mode frequencies and amplitudes}
Results from peak-bagging are displayed in Table \ref{tab_1} and represented in Fig. \ref{fig_ech} for frequencies (\'echelle diagram), and Fig. \ref{fig_W_amp} for widths and amplitudes. Frequencies  are very consistent among fitters and with VIRGO and BiSON. Except for a few peaks with low SNR, all fit within 1~$\sigma$.
In regards to mode widths, the dispersion is relatively large between fitters, with commonly a factor of two difference, but the error bars are large and mostly overlap. The measured widths from K2 are generally larger than those measured by Gaulme on simultaneous VIRGO data, but agree relatively well with those retrieved from 14 years of VIRGO/SPM (green) by Stahn (2010\footnote{https://www.mps.mpg.de/phd/theses/analysis-of-time-series-of-solar-like-oscillations-applications-to-the-sun-and-hd52265}) and simultaneous BiSON measurements.

As for the amplitudes, there is few dispersion between fitters  -- Benomar provides the lowest and Davies the largest -- but error bars mostly overlap, except for one peak at $3303\ \mu$Hz. The average amplitudes over the eight orders in common for all fitters match within $1~\sigma$ ($2.11\pm0.19$ ppm for Benomar and $2.34\pm0.15$ ppm for Davies).  Mean VIRGO amplitudes measured by Gaulme ($2.55 \pm 0.07$ ppm) are larger but still compatible with K2. However, it is obvious from Fig. \ref{fig_W_amp} that K2's amplitudes are lower than VIRGO's, especially around $\nu\ind{max}$, where VIRGO amplitudes are $\approx3.2$ ppm and K2 $\approx2.2$ ppm, i.e. 1/3 larger. This is presumably due to \textit{Kepler}'s broader and, in particular, redder passband. \citet{Jimenez_1999} showed the ratio of the mode amplitudes measured from VIRGO data for the three channels are: blue-to-green $\approx1.4$ and green-to-red $\approx 2$, which is consistent with the discrepancies we observe with respect to VIRGO green channel data. Note that BiSON amplitudes are not directly comparable because it is a velocity measurement.

\section{Conclusion}
We report the first non-direct detection of solar oscillations from intensity measurements. The use of K2 photometry of reflected light from Neptune provides sufficient SNR to detect at least eight orders with degrees $\ell=0,1,2$. 
We obtain a determination of $\Delta\nu$ that is consistent with measurements from SOHO/VIRGO/SPM and BiSON. Differences of about 2 to 3 $\sigma$, depending on methods, were observed in the determination of $\nu\ind{max}$ relative to the usual solar reference \citep[$3100\ \mu$Hz, e.g. ][]{Broomhall_2009}. The application of asteroseismic scaling relations produces a mass and radius of $1.14\pm0.06\ M_\odot$ and $1.04\pm0.02\ R_\odot$ for the Sun. 
However, a close look at the simultaneous photometric VIRGO/SPM data indicates that  $\nu\ind{max}$ was actually larger than the usual solar reference, which is not surprising given the stochastic nature of solar oscillations. By taking into account the SNR, the value of $\nu\ind{max}$ we measure from K2 data is consistent with VIRGO within 1 $\sigma$, and corresponds to the upper 20\,\% of the posterior density probability.
The peak-bagging technique leads to mean amplitude and width that match those from VIRGO within error bars. However, amplitudes are systematically lower in K2 data, by about 1/3 around $\nu\ind{max}$, which is due to the redder passband of \textit{Kepler} observations. 

\clearpage
\newpage
\begin{deluxetable*}{l l l l l l l l l l l l l l l l l}
\tabletypesize{\scriptsize}
\tablecaption{Global helioseismic parameters and results for individual peaks from K2 Neptune observations. The quantities $n$ and $\ell$ indicate oscillation mode radial orders and degrees, $\nu\ind{n,\ell}$ frequencies, $W\ind{n,\ell}$ widths, and $A\ind{n,\ell}$ amplitudes. All frequencies are expressed in $\mu$Hz, and amplitudes in ppm. \label{tab_1}}
\tablewidth{0pt}
\tablehead{
}
\startdata
\multicolumn{17}{c}{\sf\normalsize{Global Helioseismic Parameters}}\\ 
\hline 
 & &  & \multicolumn{2}{c}{Benomar} & \multicolumn{2}{c}{Corsaro} & \multicolumn{2}{c}{Davies} & \multicolumn{2}{c}{Gaulme} & \multicolumn{2}{c}{Garcia-Mathur} & \multicolumn{2}{c}{Huber} & \multicolumn{2}{c}{Mosser} \\ 
& \multicolumn{2}{l}{$\nu\ind{max}$ ($\mu$Hz)}   &  \multicolumn{2}{c}{ 3211(46)} & \multicolumn{2}{c}{3262(21)} & \multicolumn{2}{c}{3207(49)} & \multicolumn{2}{c}{3217(50)} & \multicolumn{2}{c}{3268(56) } & \multicolumn{2}{c}{3235(78)}& \multicolumn{2}{c}{3267(45)}\\ 
& \multicolumn{2}{l}{$\Delta\nu$ ($\mu$Hz)}   &  \multicolumn{2}{c}{ 134.9(1)} & \multicolumn{2}{c}{134.77(5)} & \multicolumn{2}{c}{134.6(3)} & \multicolumn{2}{c}{134.9(3)} & \multicolumn{2}{c}{135(2) } & \multicolumn{2}{c}{134.9(8)}& \multicolumn{2}{c}{135.3(3)} \\ 
& \multicolumn{2}{l}{$M\ind{ast}/M_\odot$}    &  \multicolumn{2}{c}{ 1.11(5)} & \multicolumn{2}{c}{1.17(2)} & \multicolumn{2}{c}{1.12(5)} & \multicolumn{2}{c}{1.12(5)} & \multicolumn{2}{c}{1.16(9) } & \multicolumn{2}{c}{1.14(9)}& \multicolumn{2}{c}{1.16(5)} \\ 
& \multicolumn{2}{l}{$R\ind{ast}/R_\odot$}   &  \multicolumn{2}{c}{ 1.04(2)} & \multicolumn{2}{c}{1.054(7)} & \multicolumn{2}{c}{1.04(2)} & \multicolumn{2}{c}{1.04(2)} & \multicolumn{2}{c}{1.05(3) } & \multicolumn{2}{c}{1.04(3)}& \multicolumn{2}{c}{1.05(2)} \\ 
\hline 
\multicolumn{17}{c}{\sf\normalsize{Mode fitting}}\\
\hline 
 & & \multicolumn{3}{c}{Benomar} && \multicolumn{3}{c}{Corsaro} && \multicolumn{3}{c}{Davies} && \multicolumn{3}{c}{Gaulme}  \\ 
 \cline{3-5} \cline{7-9} \cline{11-13} \cline{15-17} 
 $n$ & $\ell$ & $\nu\ind{n,\ell}$ & $W\ind{n,\ell}$ & $A\ind{n,\ell}$ && $\nu\ind{n,\ell}$ & $W\ind{n,\ell}$ & $A\ind{n,\ell}$ && $\nu\ind{n,\ell}$ & $W\ind{n,\ell}$ & $A\ind{n,\ell}$ && $\nu\ind{n,\ell}$ & $W\ind{n,\ell}$ & $A\ind{n,\ell}$ \\ 
       &         & $\mu$Hz           & $\mu$Hz           &     ppm            &&    $\mu$Hz          &   $\mu$Hz        &    ppm             && $\mu$Hz            &  $\mu$Hz         &   ppm             &&     $\mu$Hz         &   $\mu$Hz        &    ppm \\ 
\hline 
15 & 1 & \nodata & \nodata & \nodata & & 2292.2(1) & 1.3(3) & 1.9(2) & & \nodata & \nodata & \nodata & & \nodata & \nodata & \nodata \\ 
15 & 2 & \nodata & \nodata & \nodata & & 2349.9(6) & 1.5(3) & 1.1(2) & & \nodata & \nodata & \nodata & & \nodata & \nodata & \nodata \\ 
16 & 0 & \nodata & \nodata & \nodata & & 2362.7(4) & 2.3(4) & 1.8(2) & & \nodata & \nodata & \nodata & & \nodata & \nodata & \nodata \\ 
16 & 1 & \nodata & \nodata & \nodata & & 2426.5(6) & 1.6(3) & 1.3(2) & & \nodata & \nodata & \nodata & & \nodata & \nodata & \nodata \\ 
16 & 2 & 2485(2) & 1.1$_{-0.8}^{+2.0}$ & 1.1$_{-0.5}^{+0.4}$ & & 2484.5(5) & 2.8(5) & 1.6(2) & & 2485(5) & \nodata & \nodata & & 2484.4(7) & \nodata & \nodata \\ 
17 & 0 & 2494(2) & 1.1$_{-0.8}^{+2.0}$ & 1.6$_{-0.7}^{+0.6}$ & & 2494.1(7) & 2.8(5) & 2.2(2) & & 2494(3) & 7(7) & 1.5(6) & & 2494.8(9) & 2.5$_{-0.9}^{+1.5}$ & 1.9$_{-0.3}^{+0.4}$ \\ 
17 & 1 & 2559(2) & 1.1$_{-0.8}^{+2.0}$ & 1.9$_{-0.9}^{+0.7}$ & & 2559.6(4) & 2.2(4) & 1.8(2) & & 2559(2) & \nodata & \nodata & & 2559.4(6) & \nodata & \nodata \\ 
17 & 2 & 2619(1) & 1.1$_{-0.8}^{+1.8}$ & 1.3$_{-0.4}^{+0.4}$ & & 2618.8(6) & 5.4(8) & 1.8(2) & & 2620(5) & \nodata & \nodata & & 2618.9(8) & \nodata & \nodata \\ 
18 & 0 & 2629.1(8) & 1.1$_{-0.8}^{+1.8}$ & 1.7$_{-0.6}^{+0.5}$ & & 2629.6(5) & 3.6(7) & 2.0(2) & & 2629(2) & 3(3) & 2.2(4) & & 2629.1(7) & 2.6$_{-0.9}^{+1.3}$ & 2.2$_{-0.3}^{+0.4}$ \\ 
18 & 1 & 2694(1) & 1.1$_{-0.8}^{+1.8}$ & 2.1$_{-0.7}^{+0.7}$ & & 2692.0(5) & 4.3(8) & 2.9(2) & & 2692(1) & \nodata & \nodata & & 2692.4(9) & \nodata & \nodata \\ 
18 & 2 & 2754.7(9) & 0.3$_{-0.2}^{+0.9}$ & 1.2$_{-0.5}^{+0.4}$ & & 2755.6(5) & 2.2(4) & 1.9(2) & & 2755(5) & \nodata & \nodata & & 2754.9(2) & \nodata & \nodata \\ 
19 & 0 & 2764.1(8) & 0.3$_{-0.2}^{+0.9}$ & 1.7$_{-0.7}^{+0.6}$ & & 2764.2(3) & 1.6(3) & 1.3(1) & & 2764(1) & 0.9(9) & 2.4(4) & & 2763.6(3) & 0.5$_{-0.2}^{+0.3}$ & 2.0$_{-0.3}^{+0.4}$ \\ 
19 & 1 & 2828.6(3) & 0.3$_{-0.2}^{+0.9}$ & 2.0$_{-0.8}^{+0.8}$ & & 2828.5(2) & 1.7(3) & 3.0(2) & & 2828.4(5) & \nodata & \nodata & & 2828.5(1) & \nodata & \nodata \\ 
19 & 2 & 2889.2(8) & 0.7$_{-0.3}^{+0.4}$ & 1.7$_{-0.3}^{+0.3}$ & & 2889.2(3) & 2.5(3) & 1.9(1) & & 2889(5) & \nodata & \nodata & & 2888.9(2) & \nodata & \nodata \\ 
20 & 0 & 2899.6(2) & 0.7$_{-0.3}^{+0.4}$ & 2.4$_{-0.5}^{+0.4}$ & & 2899.7(1) & 1.2(2) & 2.3(2) & & 2899.6(3) & 1(1) & 2.9(4) & & 2899.7(2) & 0.6$_{-0.2}^{+0.3}$ & 2.5$_{-0.3}^{+0.4}$ \\ 
20 & 1 & 2963.8(4) & 0.7$_{-0.3}^{+0.4}$ & 2.9$_{-0.5}^{+0.5}$ & & 2963.7(1) & 1.7(2) & 3.4(2) & & 2963.6(4) & \nodata & \nodata & & 2963.9(1) & \nodata & \nodata \\ 
20 & 2 & 3025(1) & 0.8$_{-0.4}^{+0.6}$ & 1.4$_{-0.3}^{+0.3}$ & & 3025.4(3) & 1.3(2) & 1.8(2) & & 3025(5) & \nodata & \nodata & & 3024.2(4) & \nodata & \nodata \\ 
21 & 0 & 3034.1(3) & 0.8$_{-0.4}^{+0.6}$ & 1.9$_{-0.4}^{+0.5}$ & & 3034.1(2) & 1.9(3) & 1.8(1) & & 3034.1(6) & 1(1) & 2.3(4) & & 3034.1(3) & 0.9$_{-0.4}^{+0.6}$ & 2.1$_{-0.3}^{+0.4}$ \\ 
21 & 1 & 3098.8(3) & 0.8$_{-0.4}^{+0.6}$ & 2.4$_{-0.5}^{+0.5}$ & & 3099.0(2) & 2.4(3) & 2.9(2) & & 3099.0(5) & \nodata & \nodata & & 3098.9(2) & \nodata & \nodata \\ 
21 & 2 & 3159.7(9) & 1.6$_{-0.8}^{+1.3}$ & 2.0$_{-0.3}^{+0.3}$ & & 3159.1(5) & 3.0(5) & 1.9(3) & & 3160(5) & \nodata & \nodata & & 3158.9(5) & \nodata & \nodata \\ 
22 & 0 & 3168.9(6) & 1.6$_{-0.8}^{+1.3}$ & 2.7$_{-0.5}^{+0.5}$ & & 3169.4(5) & 2.4(5) & 2.4(3) & & 3169.0(8) & 2(2) & 2.8(4) & & 3168.9(7) & 1.6$_{-0.6}^{+1.0}$ & 2.8$_{-0.3}^{+0.4}$ \\ 
22 & 1 & 3233.5(4) & 1.6$_{-0.8}^{+1.3}$ & 3.3$_{-0.6}^{+0.6}$ & & 3233.5(1) & 1.7(3) & 3.9(3) & & 3233.5(4) & \nodata & \nodata & & 3233.5(3) & \nodata & \nodata \\ 
22 & 2 & 3296(1) & 2.7$_{-1.5}^{+2.0}$ & 2.4$_{-0.3}^{+0.3}$ & & 3296.7(5) & 2.4(4) & 2.6(3) & & 3296(4) & \nodata & \nodata & & 3294.8(3) & \nodata & \nodata \\ 
23 & 0 & 3303.8(4) & 2.7$_{-1.5}^{+2.0}$ & 3.3$_{-0.5}^{+0.5}$ & & 3303.8(1) & 1.2(3) & 3.6(3) & & 3304(1) & 0.9(9) & 2.6(4) & & 3303.8(2) & 1.0$_{-0.5}^{+1.0}$ & 2.5$_{-0.4}^{+0.5}$ \\ 
23 & 1 & 3368.9(8) & 2.7$_{-1.5}^{+2.0}$ & 4.0$_{-0.6}^{+0.6}$ & & 3367.1(10) & 4.4(9) & 2.2(2) & & 3368(1) & \nodata & \nodata & & 3368.3(6) & \nodata & \nodata \\ 
23 & 2 & 3431(2) & 1.3$_{-0.9}^{+2.3}$ & 1.1$_{-0.4}^{+0.4}$ & & 3429.8(8) & 6(1) & 2.7(3) & & 3431(3) & \nodata & \nodata & & 3428.5(8) & \nodata & \nodata \\ 
24 & 0 & 3439(1) & 1.3$_{-0.9}^{+2.3}$ & 1.5$_{-0.5}^{+0.5}$ & & 3440.0(7) & 3.3(8) & 1.5(2) & & 3438(3) & 2(2) & 2.0(4) & & 3439.2(6) & 2.0$_{-0.8}^{+1.3}$ & 2.1$_{-0.3}^{+0.4}$ \\ 
24 & 1 & 3505.0(9) & 1.3$_{-0.9}^{+2.3}$ & 1.9$_{-0.6}^{+0.6}$ & & 3505.3(3) & 2.4(4) & 2.5(2) & & 3505(2) & \nodata & \nodata & & 3505.2(5) & \nodata & \nodata \\ 
24 & 2 & \nodata & \nodata & \nodata & & 3566.2(7) & 5(1) & 1.7(2) & & \nodata & \nodata & \nodata & & \nodata & \nodata & \nodata \\ 
25 & 0 & \nodata & \nodata & \nodata & & 3572.1(5) & 3.0(6) & 1.6(2) & & \nodata & \nodata & \nodata & & \nodata & \nodata & \nodata \\ 
\hline 
\multicolumn{2}{c}{Mean}  & \nodata & 1(1) & 2.1(5) & & \nodata & 2.6(5) & 2.2(2) & & \nodata & 2(3) & 2.3(4) & & \nodata & 1.5(8) & 2.3(4) \\ 
\multicolumn{2}{c}{$\langle\nu_{\ell=0,1}\rangle$}  & 2999(321) & \nodata & \nodata  & & 2932(399) & \nodata & \nodata  & & 2999(321) & \nodata & \nodata  & & 2999(321) & \nodata & \nodata \\ 
\multicolumn{2}{c}{$\langle\Delta\nu_{\ell=0}\rangle$}  & 134.9(4) & \nodata & \nodata  & & 134(2) & \nodata & \nodata  & & 134.9(4) & \nodata & \nodata  & & 134.9(6) & \nodata & \nodata \\ 
\multicolumn{2}{c}{$\langle\Delta\nu_{\ell=1}\rangle$}  & 135.1(5) & \nodata & \nodata  & & 135(2) & \nodata & \nodata  & & 135(1) & \nodata & \nodata  & & 135(1) & \nodata & \nodata 
\enddata
\end{deluxetable*}


\acknowledgements{TA, PB and RAG acknowledge the support received from the CNES GOLF grant. EC and RAG received funding from the European Community's Seventh Framework Programme ([FP7/2007-2013]) under grant agreement No. 312844 (SPACEINN). E.C. has received fundings from the European Union's Horizon 2020 research and innovation programme under the Marie Sklodowska-Curie grant agreement n$^\circ$ 664931. D.H. acknowledges support by the Australian Research Council's Discovery Projects funding scheme (project number DE140101364) and support by the National Aeronautics and Space Administration under Grant NNX14AB92G issued through the Kepler Participating Scientist Program. SM would like to acknowledge support from NASA grants NNX12AE17G and NNX15AF13G and NSF grant AST-1411685.}


\end{document}